\documentclass[prl,
preprintnumbers,amsmath,amssymb,pagenumbers]{revtex4}
\usepackage{graphicx}

\relax

\def\bes{\begin{eqnarray}}
 \def\ees{\end{eqnarray}}
\def\be{\begin{equation}}
\def\ee{\end{equation}}
\def\bs{\begin{subequations}}
\def\es{\end{subequations}}
\newcommand{\een}{\end{subequations}}
\newcommand{\ben}{\begin{subequations}}
\newcommand{\beq}{\begin{eqalignno}}
\newcommand{\eeq}{\end{eqalignno}}

\def\tit{{\tilde{t}}}
 \def\ex{\epsilon}
 \def\lx{\lambda}
\def\et{{\tilde{\ex}}}
\def\lt{{\tilde{\lx}}}
 
\def\rt{{\tilde{r}}}

 \def\Ac{{\mathcal A}}
 \def\Bc{{\mathcal B}}
 \def\Cc{{\mathcal C}}
\def\Dc{{\mathcal D}}

\def \lta {\mathrel{\vcenter
     {\hbox{$<$}\nointerlineskip\hbox{$\sim$}}}}

\begin{document}

\title{On the dynamics of classicalization}

\author{N. Brouzakis$^{(1)}$, J. Rizos$^{(2)}$ and N. Tetradis$^{(1)}$}
\affiliation{
(1)
Department of Physics,
University of Athens,
University Campus,
Zographou 15784, Greece
\\
(2)
Theory Division,
Department of Physics,
University of Ioannina,
Ioannina 45110, Greece}

\begin{abstract}
We discuss the mechanism through which classicalization may occur during the collapse of a spherical field
configuration modelled as a wavepacket. We demonstrate that the phenomenon is associated with the dynamical
change of the equation of motion from a second-order partial differential equation of hyperbolic to one of elliptic type.
Within this approach, we rederive the known expression for the classicalization radius. We also find
indications that classicalization is associated with the absence of wave propagation at distances below
the classicalization radius and the generation of shock fronts. The full quantitative
picture can be obtained
only through the numerical integration of a partial differential equation of mixed type.
\end{abstract}

\maketitle

\section{Introduction}\label{intro}

The classicalization scenario advocates that the high-energy behavior of certain classes of seemingly simple
scalar field theories can be nontirivial. Typically, such theories are described by
Lagrangians which include non-renormalizable higher-derivative terms. It is possible
that scattering in the context of such theories can take place
at distances much larger than the typical length scale $L_*$ associated with the couplings multiplying the
non-renormalizable terms. The scattering scale $r_*$, which is termed classicalization radius and
determines the cross-section, depends on the center of mass energy in a nontrivial way.
The conjecture put forward in ref. \cite{dvalex1,dvpirts,dvali,dvalex2} is that distances much smaller than $r_*$ cannot
be probed, as all significant scattering takes place already at $r_*$. Thus, the loss of unitarity expected
when probing length scales $\sim L_*$, as a result of the presence of the non-renormalizable terms, never occurs.

The classicalization picture relies heavily on the analogy with black-hole formation
at ultra-Planckian energies. The high-energy scale $L_*$ is the analogue of the Planck scale, while the
classicalization radius $r_*$ the analogue of the Schwarzschild radius. The classicalon is
a configuration of the scalar field that mimicks the behavior of the black hole.
Both objects are one-parameter solutions of the equations of motion with a singularity at the origin.
In the case of the black hole the singularity is hidden behind the horizon. For the classicalon, it is usually
attributed to a source term generated
by the self-sourcing of the field in situations in which its energy is concentrated in
small regions of space. In specific examples, the field configuration is split into a part $\phi_0(t,\vec{x})$, that corresponds
to a solution of the equation of motion in the absence of the non-renormalizable terms, and a perturbation
$\phi_1(t,\vec{x})$. Perturbation theory generates an equation of motion for $\phi_1$, which includes a
source term resulting from the variation of the non-renormalizable terms evaluated for $\phi_0$.

It seems clear that the perturbative arguments need to be reinforced by a non-perturbative analysis.
As the essence of the classicalization picture is that only classical physics is relevant for the theories in
question, the necessary procedure is obvious: One needs to solve the classical equation of motion of the full theory for the
scattering problem. The purpose of this letter is to study the form of this equation for the case of a collapsing spherical
wavepacket discussed in ref. \cite{dvpirts,dvali}. Even though an accurate numerical solution is a difficult technical task that
we postpone for the future, the
general properties of the equation can be established more easily. In particular, the emergence of the classicalization radius $r_*$,
the formation of shock fronts and the occurence of scattering already at this scale,
as well as the suppression of wave propagation at length scales smaller than $r_*$,
seem plausible predictions of the equation of motion. Our analysis does not rely on the
emergence of a configuration identified with the classicalon. Nevertheless it supports the picture of scattering
at length scales much larger than the short-distance scale $L_*$.

\section{Equation of motion}\label{eomm}

We consider the theory of a scalar field $\phi(t,\vec{x}$) with derivative self-interactions. The Lagrangian density is
\be
{\cal L}=\frac{1}{2}\left(\partial_\mu\phi \right)^2-\delta\frac{L^4_*}{4}\left( \left(\partial_\mu\phi \right)^2 \right)^2,
\label{lagrangian} \ee
with the Minkowski metric given by $\eta_{\mu\nu}={\rm diag} (1,-1,-1,-1)$. We allow for both signs of the higher-derivative
term by considering the values $\delta=\pm1$.
The equation of motion of the field $\phi$ is
\be
\partial^\mu\left[  \partial_\mu \phi \left( 1- \delta L^4_*\left(\partial_\nu\phi \right)^2\right) \right]=0.
\label{eom} \ee
We are interested in a configuration described by a collapsing spherical wavepacket represented by a Gaussian
of width $a$ initially centered around a radius $r_0$. It has the form
\be
\phi_0(t,r)=\frac{A}{r}\exp\left[-\frac{\left(r+t-r_0\right)^2}{a^2} \right].
\label{wave} \ee
This configuration is an exact solution of the equation of motion in the absence of the higher-derivative term ($L_*=0$).
It is also an approximate solution for nonzero $L_*$ and sufficiently large values of $r_0$.

When expressed in spherical coordinates, eq. (\ref{eom}) assumes the form
\be
\left(1-3 \lx \phi_t^2+\lx\phi_r^2 \right)\phi_{tt}-\left(1- \lx \phi_t^2+3\lx\phi_r^2 \right)\phi_{rr}
+4\lx \phi_r\phi_t \,\phi_{tr}=\frac{2\phi_r}{r}\left(1- \lx \phi_t^2+\lx\phi_r^2 \right),
\label{eomsph} \ee
where $\lx=\delta L^4_*$ and subscripts denote partial derivatives. The above equation can be
expressed as a conservation law, in the form
\be
\partial_t\left[\phi_t \left(1- \lx \phi_t^2+\lx\phi_r^2 \right)\right]
-\frac{1}{r^2}\partial_r\left[r^2\phi_r \left(1- \lx \phi_t^2+\lx\phi_r^2 \right)\right]
=0.
\label{eomcons} \ee
The initial conditions for its solution are of the Cauchy type:
\be
\phi(0,r)=\phi_0(0,r),
~~~~~~~~~~~~~~~~~~~~~~~
\partial_t\phi(0,r)=\partial_t \phi_0(0,r),
\label{init} \ee
where we assume that $r_0$ is much larger than any other physical scale.

In refs. \cite{dvpirts,dvali} eq. (\ref{eomsph}) is solved by writing $\phi(t,r)=\phi_0(t,r)+\phi_1(t,r)$, with $\phi_0$ given by
eq. (\ref{wave}). The correction
$\phi_1$ is treated as a perturbation that vanishes for $L_*= 0$.  It is shown that $\phi_1$ becomes comparable to $\phi_0$ at the
classicalization radius
\be
r_*\sim L_* \left( \frac{A^2 L_*}{a}\right)^{1/3}.
\label{clr} \ee
The subsequent evolution of the field cannot be derived reliably within perturbation theory. Moreover, it is not
clear whether the initial configuration, despite its deformation, continues its propagation towards smaller values of $r$, or
whether there is significant scattering towards large $r$.

The properties of eq. (\ref{eomsph}) become more transparent if it is written in the form of a quasi-linear second-order partial
differential equation:
\be
\Ac(\phi_t,\phi_r) \, \phi_{tt}+ \Bc(\phi_t,\phi_r) \, \phi_{tr}+ \Cc(\phi_t,\phi_r) \, \phi_{rr}= \Dc(\phi_t,\phi_r,r),
\label{pdeform} \ee
with
\begin{eqnarray}
\Ac(\phi_t,\phi_r)&=&1-3 \lx \phi_t^2+\lx\phi_r^2
\label{aa} \\
\Bc(\phi_t,\phi_r)&=&4\lx\phi_t\phi_r
\label{bb} \\
\Cc(\phi_t,\phi_r)&=&-\left(1- \lx \phi_t^2+3\lx\phi_r^2\right)
\label{cc} \\
\Dc(\phi_t,\phi_r,r)&=&\frac{2\phi_r}{r}\left(1- \lx \phi_t^2+\lx\phi_r^2 \right).
\label{dd} \end{eqnarray}
The type of this partial differential equation is determined by the discriminant
\be
\Delta=\Bc^2-4\Ac\Cc=12\left(\frac{1}{3}-\lx \phi^2_t+\lx \phi_r^2 \right)\left(1-\lx \phi^2_t+\lx \phi_r^2 \right).
\label{discr} \ee
For $\Delta>0$ the equation is hyperbolic,
for $\Delta=0$ parabolic, while for $\Delta<0$ elliptic.
Under an appropriate coordinate transformation $(t,r)\to(\alpha,\beta)$, a hyperbolic partial differential equation can be
written in the form $\phi_{\alpha\alpha}-\phi_{\beta\beta}+\, ...=0$, while an elliptic one as
 $\phi_{\alpha\alpha}+\phi_{\beta\beta}+\, ...=0$, with the dots denoting lower-derivative terms.
Hyperbolic equations have wave-like solutions, while elliptic ones do not
support propagating solutions. It is clear that, depending on the form of $\phi(t,r)$, eq. (\ref{pdeform}) can be of any type. Moreover,
it is possible that it may change type dynamically during the time evolution of an initial configuration. Equations that have this property are
characterized as of mixed type and are notoriously difficult to integrate.

\begin{figure}[t]
\includegraphics[width=160mm,height=90mm]{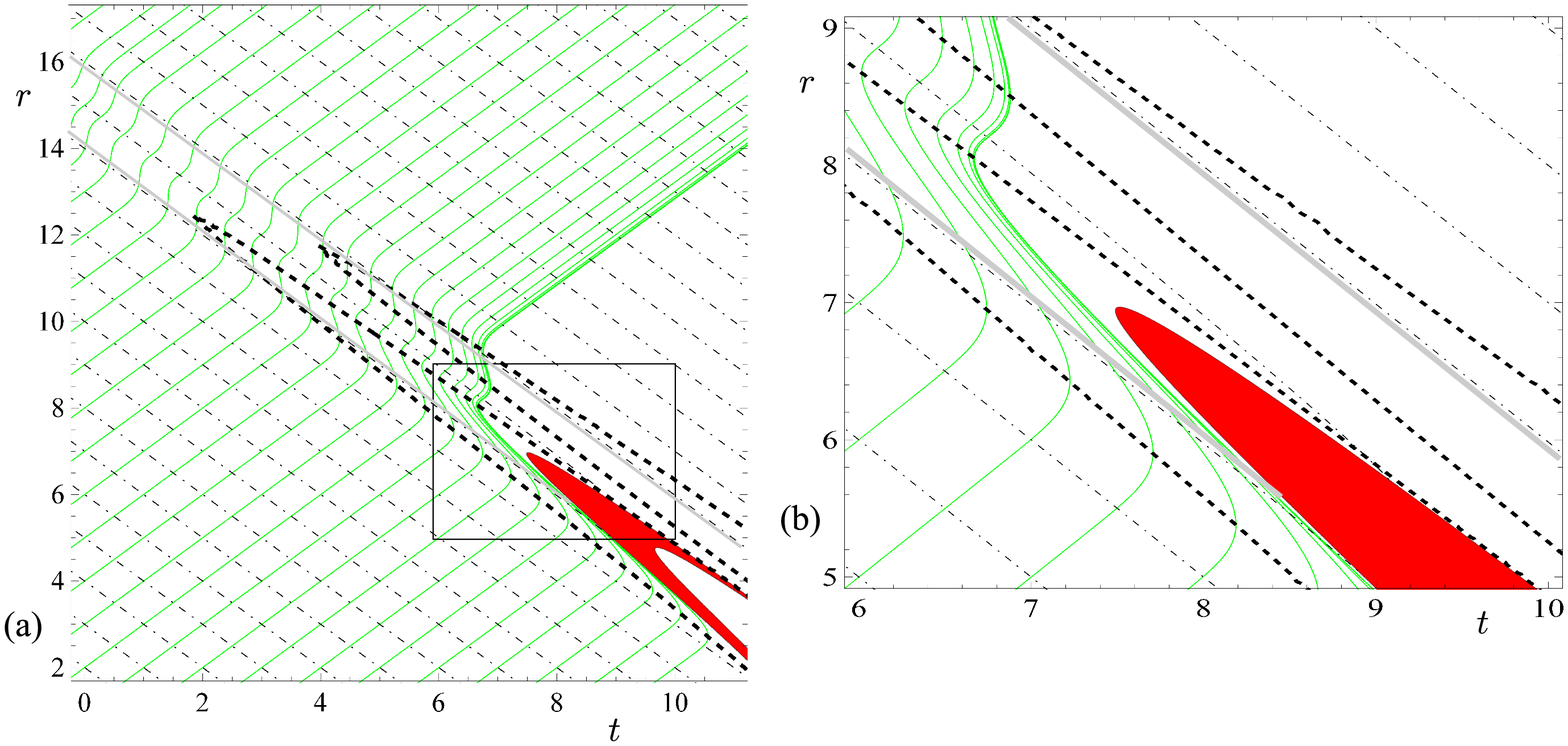}
\caption{The characteristics of the field equation (\ref{pdeform}). The functions
$\Ac(\phi_t,\phi_r)$, $\Bc(\phi_t,\phi_r)$, $\Cc(\phi_t,\phi_r)$ are
evaluated for the configuration $\phi_0(r,t)$ of eq. (\ref{wave}) , with  $a=1$, $A=10$, $r_0=15$,
in units such that $L_*=1$, $\lx=1$.
The equation is of the elliptic type within the shaded (red) area. The panel (b) is a magnification of the
area denoted by a square in panel (a).}
\label{fig}
\end{figure}

For large $r$ we have
$\Ac(\phi_{0t},\phi_{0r})\simeq\Cc(\phi_{0t},\phi_{0r})\simeq 1$ and $\Bc(\phi_{0t},\phi_{0r})\simeq 0$, where
$\Ac$, $\Bc$, $\Cc$ have been evaluated for the configuration $\phi_0(t,r)$ of eq. (\ref{wave}).
In this limit, $\Delta \simeq 4$ and eq. (\ref{pdeform}) becomes the wave equation in spherical coordinates. The configuration
(\ref{wave}) is an approximate solution, as it is a superposition of spherical waves.
It is possible however, that the subsequent evolution of $\phi(t,r)$ may result in the change of the type of eq. (\ref{pdeform}).
We demonstrate this possiblity by depicting in fig. \ref{fig} the region in which $\Delta <0$, when $\Ac$, $\Bc$, $\Cc$ are
evaluated for the configuration (\ref{wave}) with $a=1$, $A=10$, $r_0=15$. We consider the case with $\delta=1$ and
work in units such that
$L_*=1$ and $\lx=1$. The region in which eq. (\ref{pdeform}) becomes elliptic corresponds to the shaded (red) area on the
$(t,r)$ plane. It must be emphasized that eq. (\ref{wave}) is not a solution of eq. (\ref{pdeform})
for late times and small values of $r$, so that the change of equation type must be viewed only as a possibility.
The derivation of an accurate (numerical) solution of eq. (\ref{pdeform}) with initial conditions given by
(\ref{init}) is necessary in order to obtain a definite confirmation of this conclusion.

It is interesting to estimate the distance $r$ at which eq. (\ref{pdeform}) changes type if $\Ac$, $\Bc$, $\Cc$ are
evaluated for the configuration (\ref{wave}). The discriminant vanishes for
\be
\rt^4=-k \lt A^2 e^{-2\et^2} \left(1+4 \rt \et \right),
\label{disc0} \ee
where $\rt=r/a$, $\rt_0=r_0/a$, $\lt=\lx/a^4$, $\et=\rt+\tit-\rt_0$ and $k$ takes the values 1 or 3 (corresponding to the
inner and outer curves bounding the shaded (red) region in fig. \ref{fig}).
The relevant values of the parameter $\et$ are of order 1, as the Gaussian falls off very rapidly outside this range.
We are interested in the region $\rt \gg 1$, because we expect the classicalization radius to be much larger than the width of
the Gaussian. A solution can be obtained for $\et \simeq -1$ and $\rt |\et| \gg 1$.
We obtain
\be
\rt_* \simeq \left(\lt A^2 \right)^{1/3},
\label{estim} \ee
an expression which coincides with the definition of the classicalization radius $r_*$ in eq. (\ref{clr}).
It is clear that $r_*$ can be interpreted as the radius at which the partial differential equation (\ref{pdeform})
 is expected to switch from hyperbolic to elliptic. The change does not occur simultaneously at all values of $r$.
For example, the center of the Gaussian corresponds to $\et\simeq 0$,
so that eq. (\ref{disc0}) is never satisfied for $\lx >0$.
This indicates that the equation changes type first at the front of the wavepacket, i.e.
for values of $r$ smaller than the one corresponding to the center of the Gaussian.

\begin{figure}[t]
\includegraphics[width=160mm,height=90mm]{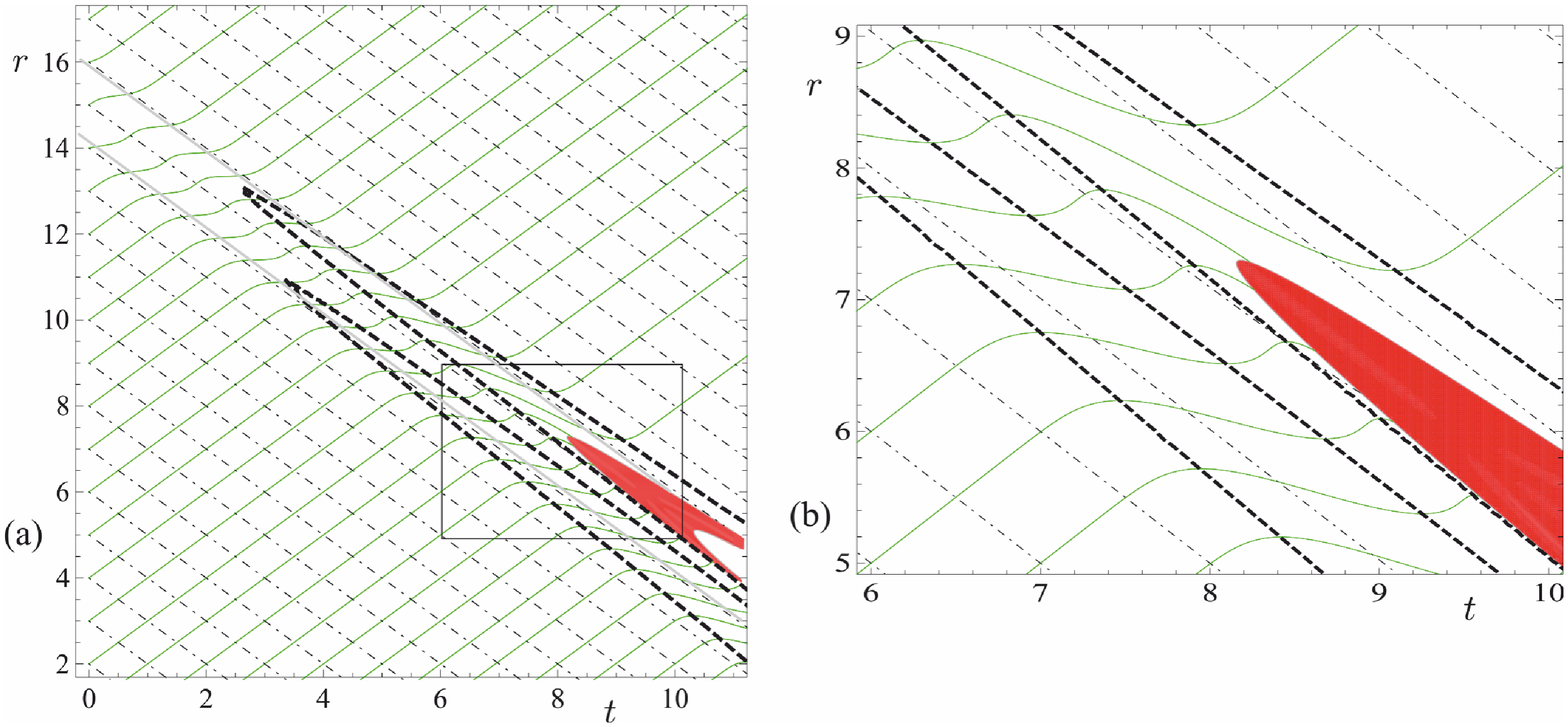}
\caption{The characteristics of the field equation (\ref{pdeform}) for $a=1$, $A=10$, $r_0=15$ and $\lx=-1$.
The panel (b) is a magnification of the area denoted by a square in panel (a).}
\label{figex}
\end{figure}

The analysis can be repeated for $\delta=-1$, which implies $\lx<0$ and corresponds to a higher-derivative term with
positive sign in the Lagrangian (\ref{lagrangian}).
The change of type of the equation of motion also occurs in this case. In fig. \ref{figex} we depict the
region in which $\Delta <0$, when $\Ac$, $\Bc$, $\Cc$ are
evaluated for the configuration (\ref{wave}) with $a=1$, $A=10$, $r_0=15$ and $\lx=-1$.
The region in which eq. (\ref{pdeform}) becomes elliptic corresponds to the shaded (red) area on the
$(t,r)$ plane. This area is similar to the one depicted in fig. \ref{fig}, but slightly displaced. The
analysis of the previous paragraph, leading to the estimate of the classicalization radius, can be repeated for
$\lx<0$. The result is
\be
\rt_* \simeq \left(|\lt| A^2 \right)^{1/3},
\label{estimex} \ee
which again reproduces eq. (\ref{clr}).
The main difference is that
the type change now occurs first at $\et \simeq 1$, i.e. at a point
at the rear of the wavepacket.
Also, in this case there is a solution of
eq. (\ref{disc0})  for $\et=0$, given by
$\rt_c \simeq (|\lt| A^2 )^{1/4}.$
For $\rt_*\gg 1$ we have $\rt_* > \rt_c$.

\section{Characteristics}\label{char}

In order to obtain more intuition on the nature of the solutions of eq. (\ref{pdeform}), we examine its characteristics
when $\Ac(\phi_t,\phi_r)$, $\Bc(\phi_t,\phi_r)$, $\Cc(\phi_t,\phi_r)$ are
evaluated for the configuration $\phi_0(r,t)$ of eq. (\ref{wave}). The characteristics are solutions of the equation
\be
\frac{dr}{dt}=\frac{1}{2\Ac}\left( \Bc\pm\sqrt{\Delta} \right).
\label{eqchar} \ee
Clearly, they can be defined only for $\Delta \geq 0$, and they do not exist in the region where the equation is elliptic.
The two signs correspond to two families of characteristics.
For large $r$ and early times we have $\Ac\simeq \Cc \simeq 1$, $\Bc\simeq 0$. Eq. (\ref{pdeform}) becomes the wave equation,
which has the configuration $\phi_0(t,r)$ as
an approximate solution. The characteristics correspond to the lines $r+t=c_1$, $r-t=c_2$.
The configuration $\phi_0(t,r)$ is projected only on the first family, as it is composed from ingoing spherical waves, which are
functions of $r+t$.

The deformation of the characteristics in the vicinity of the region in which the equation changes type is depicted in fig. \ref{fig}
for a model with $a=1$, $A=10$, $r_0=15$. We consider the case $\delta=1$ and use units such that $L_*=1$, $\lx=1$.
The two families correspond to solutions of eq. (\ref{eqchar}) expressed in the form
$u(t,r)=c_1$ and $v(t,r)=c_2$. Each value of $c_1$ or $c_2$ determines a curve on the $(t,r)$ plane, along which initial
disturbances propagate. We choose the first family (which we term ingoing) as the one that reduces to $u(t,r)=r+t$ at initial times and large $r$,
while the second one (termed outgoing) reduces to $v(t,r)=r-t$.
The dash-dotted (black) lines in fig. \ref{fig} depict ingoing, while the solid (green) lines outgoing characteristics.

The initial configuration $\phi_0(t,r)$ is projected only on the first family and within a limited range of $r$. For $t=0$ the initial
Gaussian is centered at $r_0=15$ and has a width $a=1$.
The initial disturbance is localized roughly in the region $14 \lta r \lta 16$. At later times it propagates along the ingoing
characteristics that start within this range. We have indicated its position by emphasizing the ingoing characteristics starting at
$r=14$ and $r=16$, which roughly bound the disturbance. The ingoing characteristics remain linear to a good approximation,
even in the vicinity of the shaded (red) region.
They terminate on the boundary of this surface. The characteristics starting at $r<15$, which correspond to points at the front of
the wavepacket, reach the shaded region before the center of the Gaussian, which starts at $r=15$.

The outgoing characteristics are deformed strongly in the vicinity of the shaded area.
Even though they are equidistant near the $r$-axis, they come very close, almost overlapping, near the shaded area.
This indicates that distinct points of an outgoing configuration tend to merge when they pass through this area.
Another striking feature is that the ``velocity'' $dr/dt$ diverges at certain points along the characteristics. All these
points are located on the lines determined by the condition $\Ac=0$, which are depicted by dashed thick lines in
fig. \ref{fig}. The form of the outgoing characteristics is visible more clearly in the enlarged region (part (b) of fig. \ref{fig})
which corresponds to the square in part (a). This form is consistent with the appearance of
outgoing shock fronts. We have depicted only characteristics emanating from the left of the shaded (red) region in order to
demonstrate their convergence. There are also outgoing characteristics starting on the boundary of the (red) region and
covering the right part of the diagram. We do not depict those for clarity of the picture.

The form of the characteristics for a model with $a=1$, $A=10$, $r_0=15$ and $\lx=-1$ is depicted in fig. \ref{figex}.
The ingoing characteristics remain linear to a good approximation, similarly to the case with $\lx=1$.
They terminate on the boundary of the shaded (red) surface.
The characteristics starting at $r>15$, which correspond to points at the rear of
the wavepacket, reach the shaded region before the center of the Gaussian, which starts at $r=15$.
The outgoing characteristics are again deformed strongly in the vicinity of the shaded area.
Even though they are equidistant near the $r$-axis, they approach each other near the shaded area.
Distinct points of an outgoing configuration tend to merge when they pass through this area.
Again this suggests the formation of a shock front.
However, the various points subsequently move away from each other and the characteristics become dilute at large $t$.
Another striking feature is that the ``velocity'' $dr/dt$ vanishes at the points along the characteristics where $\Cc=0$.
These are depicted by dashed thick lines in fig. \ref{figex}.

\section{Conclusions}\label{concl}

Our study of the partial differential equation (\ref{pdeform}) has been based on evaluating the
coefficient functions $\Ac(\phi_t,\phi_r)$, $\Bc(\phi_t,\phi_r)$, $\Cc(\phi_t,\phi_r)$
for the spherical wavepacket $\phi_0(r,t)$ given by eq. (\ref{wave}). This configuration is an approximate solution of the exact
eq. (\ref{pdeform}) for early times and large $r$. However, it is not expected to remain so when the center of the wavepacket
approaches the region near $r=0$. The main prediction of the classicalization scenario is that the deviations become large already
at the classicalization radius $r_*$ given by eq. (\ref{clr}), and not at the fundamental scale $L_*$. We have analyzed this possibility
 by examining the nature of the solutions of eq. (\ref{pdeform}).

Our main observation is that eq. (\ref{pdeform}) is a partial differential equation of mixed type. For the initial conditions (\ref{init})
the equation is hyperbolic in the part of the $(t,r)$ plane where the field is nonvanishing initially.
At later times the field fluctuation moves into a
region where the equation becomes elliptic. Wave propagation is not supported by equations of elliptic type, and is probably replaced
by exponential decay. On the other hand, it seems also plausible that the inability of the fluctuation to propagate towards
smaller values of $r$ will be accompagnied by, at least partial, reflection. The situation is complicated by the indications for outgoing
shock-front formation that we discussed in the previous section. We expect that these outgoing shock fronts will form first at the
front or rear of the wavepacket, depending on the sign of the higher derivative term in the Lagrangian (\ref{lagrangian}).
Subsequently, they
will interact with the remaining part of the ingoing wavepacket. Clearly, our current analysis cannot provide any clues
on the details of this process. We also mention at this point that the numerical treatment of the formation of shock fronts
is a very difficult problem, for which a general framework has not been developed yet. Instead, the analysis is carried out on
a case by case basis.

Despite neglecting the effect of backreaction in the results of the previous sections,
we may attempt to draw some quantitative conclusions. The ingoing characteristics in fig. \ref{fig}
remain largely unaffected up to the
point where they terminate on the surface $\Delta =0$. As a result, we expect that the reflection
of (part of) the wavepacket will take place within a short range of values of $r$. The relevant scale
is given by the classicalization radius of eq. (\ref{clr}).
Also, the divergence of the speed of transmission observed at certain points on the outgoing characteristics is
an artefact of our approximate treatment of eq. (\ref{pdeform}). In an exact solution, causality would prevent this
divergence. However, the merging of characteristics, typical of the appearance of shock fronts,
is expected to persist.

As a final comment, we mention that the approach that we outlined is applicable also to cases in which the
phenomenon of classicalization is not expected to occur.
For example, for  planar configurations moving along the $z$-axis the equation of motion
becomes
\be
\left(1-3 \lx \phi_t^2+\lx\phi_z^2 \right)\phi_{tt}-\left(1- \lx \phi_t^2+3\lx\phi_z^2 \right)\phi_{zz}
+4\lx \phi_z\phi_t \,\phi_{tz}=0.
\label{eompl} \ee
An initial field configuration of the form
\be
\phi_0(t,z)=A\exp\left[-\frac{\left(z+t-z_0\right)^2}{a^2} \right]+A\exp\left[-\frac{\left(z-t+z_0\right)^2}{a^2} \right]
\label{plwave} \ee
describes two localized planar wavepackets moving in opposite directions. It is an approximate solution of
eq. (\ref{eompl}) as long as there is no significant overlap of the two Gaussians.
The discriminant of the partial differential equation (\ref{eompl}) is
\be
\Delta=12\left(\frac{1}{3}-\lx \phi^2_t+\lx \phi_z^2 \right)\left(1-\lx \phi^2_t+\lx \phi_z^2 \right).
\label{discrpl} \ee
Evaluating it for the configuration (\ref{plwave}) gives $\Delta\simeq 4$ (the equation is hyperbolic),
apart from the region of significant overlap of
the wavepackets. This is is a region of width $\sim a$ around the origin, appearing at times $t\sim z_0$.
The classicalization phenomenon is not expected to occur in this case, in agreement with the analysis of
ref. \cite{akhoury}.

\section*{Acknowledgments}
We would like to thank G. Dvali for useful discussions and the CERN Theory Division for hospitality during the completion of
this work.  Our research was supported in part by
the ITN network ``UNILHC'' (PITN-GA-2009-237920).

\end{document}